\documentclass[10pt]{article}
\usepackage{a4wide}                      
\usepackage{epsf}                        
\usepackage{latexsym}                    
\usepackage{amsfonts}                    

\usepackage{cite}

\newcommand{\sect}[1]{\setcounter{equation}{0}\section{#1}}
\renewcommand{\theequation}{\arabic{section}.\arabic{equation}}

\usepackage[unicode=true,colorlinks=true, urlcolor=blue, linkcolor=blue, citecolor=blue]{hyperref}
\def\be{\begin{equation}}
\def\ee{\end{equation}}
\def\bea{\begin{eqnarray}}
\def\eea{\end{eqnarray}}

\def\nnw{\nonumber \\ [.2cm]}

\def\hsp#1{\hspace*{#1}}

\def\part{\partial}
\def\tfrac#1#2{{\textstyle{\frac{#1}{#2}}}}
\def\half{\tfrac{1}{2}}

\def\cR{{\cal R}}

\def\dder{\mathrm{d}}
\def\oGamma{{\mathring \Gamma}}

\def\cL{{\cal L}}  
\def\sqrtg{\sqrt{|g|}}

\def\bGamma{\bar{\Gamma}}

\def\mn{{\mu\nu}}
\def\mnr{{\mu\nu\rho}}

\def\makeatletter{\catcode`\@=11}
\makeatletter
\def\mathbox#1{\hbox{$\m@th#1$}}%
\def\math@ccstyles#1#2#3#4#5#6#7{{\leavevmode
      \setbox0\mathbox{#6#7}%
      \setbox2\mathbox{#4#5}%
      \dimen@ #3%
      \baselineskip\z@\lineskiplimit#1\lineskip\z@
      \vbox{\ialign{##\crcr
             \hfil \kern #2\box2 \hfil\crcr
             \noalign{\kern\dimen@}%
             \hfil\box0\hfil\crcr}}}}
\def\mathaccstyles{\math@ccstyles\maxdimen}
\def\maththroughstyles{\math@ccstyles{-\maxdimen}}
\def\unity%
 {\maththroughstyles{.45\ht0}\z@\displaystyle {\mathchar"006C}\displaystyle 1}
\begin{document}


\rightline{\today}
{~}
\vspace{1.4truecm}

\centerline{\LARGE \bf A non-trivial connection for the metric-affine}
\vspace{.5truecm}
\centerline{\LARGE \bf  Gauss-Bonnet theory in $D=4$}
\vspace{1.3truecm}

\centerline{
    {\large \bf Bert Janssen,}
    {\large \bf Alejandro Jim\'enez-Cano and}
    {\large \bf Jos\'e Alberto Orejuela}\footnote{E-mail addresses: 
                                  bjanssen@ugr.es, alejandrojc@ugr.es, 
                                 josealberto@ugr.es
}                      
                                                            }
\vspace{.4cm}
\centerline{{\it Departamento de F\'{\i}sica Te\'orica y del Cosmos} and}
\centerline{{\it Centro Andaluz de F\'{\i}sica de Part\'{\i}culas Elementales}}
\centerline{{\it Facultad de Ciencias, Avda Fuentenueva s/n,}}
\centerline{{\it Universidad de Granada, 18071 Granada, Spain}}

\vspace{2truecm}

\centerline{\bf ABSTRACT}
\vspace{.5truecm}

\noindent
We study non-trivial ({\it i.e.}\ non-Levi-Civita) connections in metric-affine Lovelock
theories. First we study the projective invariance of general Lovelock actions and show
that all connections constructed by acting with a projective transformation of the Levi-Civita
connection are allowed solutions, albeit physically equivalent to Levi-Civita. We then show
that the (non-integrable) Weyl connection is also a solution for the specific case of the
four-dimensional metric-affine Gauss-Bonnet action, for arbritrary vector fields.
The existence of this solution is related to a two-vector family of transformations, that
leaves the Gauss-Bonnet action invariant when acting on metric-compatible connections.
We argue that this solution is physically inequivalent to the Levi-Civita connection, giving
thus a counterexample to the statement that the metric and the Palatini formalisms are
equivalent for Lovelock gravities. We discuss the mathematical structure of the set of
solutions within the space of connections.

\newpage

\sect{Introduction}
\label{introduction}

Metric-affine gravity (sometimes also called the Palatini formalism) is a set of theories
in which the metric $g_\mn$ and the affine connection $\Gamma_\mn{}^\rho$ are taken to be
independent variables. They are extensions of the more familiar metric theories of gravity,
which consider only the metric as a dynamical variable and presuppose invariably the affine
connection to be the Levi-Civita connection of the metric,
\be
\mathring\Gamma_\mn{}^\rho  \ = \ \half  g^{\rho\lambda}
   \Bigl(\partial_\mu g_{\lambda\nu} \ + \ \partial_\nu g_{\mu\lambda}
   \ - \ \partial_\lambda g_\mn \Bigr).
\ee
In metric-affine theories, however,  the idea is that the affine connection $\Gamma_\mn{}^\rho$
should be determined by its own equation of motion, just as any other dynamical variable
of the theory.

It has been shown \cite{Pons, BJJOSS} (see also \cite{JJOS}) that the physics described by
the Einstein-Hilbert-Palatini action,
$S=\frac{1}{2\kappa}\int \dder^Dx \sqrtg \cR(\Gamma)$ with $D>2$, possibly extended with a
minimally coupled matter Lagrangian, is equivalent to the usual metric Einstein-Hilbert
action, even though it allows a more general affine connection,
\be
\bGamma_\mn{}^\rho \ = \ {\mathring\Gamma}_\mn{}^\rho \ + \ A_\mu \,\delta_\nu^\rho,
\label{PalatiniGamma}
\ee
with $A_\mu$ an arbitrary vector field. Indeed, the vector field $A_\mu$ does not have
any physically measurable influence, as can be seen in the fact that the Einstein equation
and the geodesic equation
are identical in the metric and the Palatini formalism. However, one can adscribe a
geometrical meaning to $A_\mu$, since $A_\mu$ can be related to the reparametrisation 
freedom of geodesics \cite{BJJOSS}: affine geodesics of the connection
$\bGamma_\mn{}^\rho$ turn out to be pre-geodesics of the Levi-Civita connection
$\oGamma_\mn{}^\rho$, through the reparametrisation
\be
\frac{\dder\tau}{\dder\lambda}(\lambda) \ =  \
   \exp  \left[\int_0^\lambda  A_\rho\, \frac{\dder x^\rho}{\dder\lambda'}
   \, \dder\lambda'\right],
\ee
where $\lambda$ is the affine parameter for the $\bGamma_\mn{}^\rho$ geodesics
and $\tau$ the proper time along the Levi-Civita ones.

Both the existence of the non-trivial solution (\ref{PalatiniGamma}) and the absence of
physical meaning for $A_\mu$ can be understood as a consequence of the projective symmetry
\be
\Gamma_\mn{}^\rho \ \rightarrow \ \Gamma_\mn{}^\rho \ + \ A_\mu \,\delta_\nu^\rho
\label{projtransf}
\ee
of the
metric-affine Einstein-Hilbert action \cite{Eisenh, JS}. Indeed, the Riemann tensor
transforms under the projective transformation as
$\cR_ \mnr{}^\lambda  \rightarrow \cR_ \mnr{}^\lambda
\, +  \, 2\partial_{[\mu} A_{\nu]}  \, \delta_\rho^\lambda$,
leaving hence the Ricci scalar $\cR = g^{\mu\rho}\,\delta_\lambda^\nu\,\cR_\mnr{}^\lambda$
invariant.
Being $\oGamma_\mn{}^\rho$ a straightforward solution to the connection equation, any
connection generated by applying a projective transformation on it is also a
(physically equivalent) solution.

The equivalence between the metric and metric-affine formalism does not extend in general
to other gravitational actions (see for example \cite{CdL}). 
Often the non-equivalence of the corresponding gravitational actions is used to construct 
models of modified gravity, with new affine degrees of freedom, 
that might yield resolution of singularities, alternatives to inflation,
dark matter or dark energy \cite{CMQ, Querrella, ABFO, SL, LBM, BD, CDV, Bauer, Olmo, ORS,
OR, BCOR, BSSW, BOR, SzySta}

There is a class of metric-affine theories for which the Levi-Civita connection is guaranteed
to be a solution of the connection equation. Indeed, in \cite{ESJ, BJB, DP} it was shown that
the Levi-Civita connection is a solution for metric-affine lagrangians of the type
$\cL(g_\mn, \cR_\mnr{}^\lambda)$, only if the lagrangian is of the Lovelock type (or at least
mimics the symmetries of curvature tensors in the Lovelock lagrangian \cite{BJB}). Hence,
for Lovelock gravities, the metric formalism is a consistent truncation of the Palatini
formalism \cite{DP}.
However it is by no means clear whether for these theories both formalisms are equivalent,
as in the case of the Einstein-Hilbert action, in the sense that all allowed solutions of the
connection
equation of the metric-affine Lovelock lagrangians yield the same physics as the metric
formalism. In other words, whether the Levi-Civita connection (possibly up to a projective
transformation) is the only solution to the connection equation.

The aim of this paper is to show that in fact they are not, by presenting an
explicit counterexample of a very specific Lovelock theory, the Weyl connection in
four-dimensional metric-affine Gauss-Bonnet theory, though we
believe the result is general for any metric-affine $k$-th order Lovelock term in $D=2k$
dimensions. Note that the $k$-th order Lovelock term in $D=2k$ dimensions is a topological
term in the metric formalism \cite{Zumino}, but not necessarily for metric-affine gravity.
We will argue that the solution is physically not equivalent to the Levi-Civita connection,
which in our opinion is an indication for the non-topological character of general $D=4$
metric-affine Gauss-Bonnet theory.

The organisation of this paper is as follows: in sections \ref{EOM} and \ref{lc}
we review the
metric-affine Gauss-Bonnet term in arbitrary dimensions, study the symmetries of the action
and write down the equations of motion in a closed form. In section \ref{Wsolution}
we show that the Weyl connection is a general solution to both the equations of motion
of the metric and the connection for the four-dimensional Gauss-Bonnet action. In section
\ref{symmetry}, we relate the existence of the solution to the projective symmetry of any
Lovelock action and a vector symmetry the four-dimensional Gauss-Bonnet in the presence of
metric-compatible  connections. Finally, in section \ref{conclusions}, study the structure
of the space of solutions of the Gauss-Bonnet action and state our
conclusion.

\noindent
\sect{Metric-affine Lovelock theory}
\label{EOM}
The $D$-dimensional $k$-th order Lovelock term in the metric-affine formalism is defined as
\bea
S \ = \ \int \dder^Dx \, \sqrtg \
\delta^{\mu_1\nu_1\dots \mu_k\nu_k}_{\alpha_1\beta_1\dots \alpha_1 \beta_k} \, \,
\mathcal{R}_{\mu_1\nu_1}{}^{\alpha_1\beta_1}(\Gamma) \, \dots \,
\mathcal{R}_{\mu_k\nu_k}{}^{\alpha_k\beta_k}(\Gamma),
\label{Lovelock-Palatini}
\eea
where we used the following conventions for the Riemann tensor and the antisymmetrised
Kronecker delta,
\bea
\mathcal{R}_{\mu\nu}{}^{\rho\lambda}(\Gamma)
&=& g^{\rho\sigma} \, \mathcal{R}_{\mu\nu\sigma}{}^\lambda(\Gamma), \nnw
\mathcal{R}_{\mu\nu\sigma}{}^\lambda (\Gamma)
&=& \partial_\mu \Gamma_{\nu\sigma}{}^\lambda \ - \ \partial_\nu \Gamma_{\mu\sigma}{}^\lambda \ 
                   + \ \Gamma_{\mu\kappa}{}^\lambda \, \Gamma_{\nu\sigma}{}^\kappa
\ - \  \ \Gamma_{\nu\kappa}{}^\lambda \, \Gamma_{\mu\sigma}{}^\kappa, \\ [.2cm]
\delta^{\mu_1\nu_1\dots\mu_k\nu_k}_{\alpha_1\beta_1\dots \alpha_k\beta_k}\,
   &=& \delta^{[\mu_1}_{\alpha_1} \, \delta^{\nu_1}_{\beta_1} \dots \delta^{\mu_k}_{\alpha_k} \, \delta^{\nu_k]}_{\beta_k} \nnw
   &=& \tfrac{(-1)^{D-1}}{(2k)! (D-2k)!} \ |g| \
       \varepsilon^{\mu_1\nu_1\dots\mu_k\nu_k\sigma_1 \dots \sigma_{D-2k}}\
       \varepsilon_{\alpha_1\beta_1\dots \alpha_k\beta_k\sigma_1 \dots \sigma_{D-2k}},
       \nonumber
\eea
with $\varepsilon_{\mu_1 \dots \mu_D}$ the completely alternating Levi-Civita symbol. 

When varying the action in metric-affine gravity, it is often useful to define the tensor
\be
\Sigma^{\mu\nu\alpha}{}_\beta
\ = \ \frac{1}{\sqrtg} \, \frac{\delta S}{\delta \mathcal{R}_{\mu\nu\alpha}{}^\beta},
\label{Sigmadef}
\ee
which for the Lovelock action (\ref{Lovelock-Palatini}) is given by
\bea
\Sigma^{\mu\nu}{}_{\alpha\beta} \ = \ g_{\alpha \sigma} \, \Sigma^{\mu\nu\sigma}{}_\beta
\ = \ k \ \delta^{\mu\nu\rho_2\lambda_2 \dots \rho_k\lambda_k}_{\alpha\beta\gamma_2\epsilon_2 \dots \gamma_k\epsilon_k} \
\mathcal{R}_{\rho_2\lambda_2}{}^{\gamma_2\epsilon_2} \, \dots \,
\mathcal{R}_{\rho_k\lambda_k}{}^{\gamma_k\epsilon_k}.
\label{LL-Sigma}
\eea
Note that in general $\Sigma^{\mu\nu}{}_{\alpha\beta}$ is antisymmetric in the first pair
of indices and for Lovelock actions (\ref{Lovelock-Palatini}) also in the last pair,
but that the latter is not true in general.

In terms of $\Sigma^{\mu\nu\alpha}{}_\beta$, the $k$-th order Lovelock action
(\ref{Lovelock-Palatini}) can be written as
\be
S \ = \ \frac{1}{k} \int \dder^Dx \, \sqrtg \ \mathcal{R}_{\mu\nu\alpha}{}^\beta \,
\Sigma^{\mu\nu\alpha}{}_{\beta},
\label{Lovelock-Palatini2}
\ee
and hence the equations of motion of the metric and the
connection respectively are given by
\bea
&& \mathcal{R}_{\mu\nu\rho}{}^\lambda \, \Sigma^{\mu\nu}{}_{\sigma\lambda} \ + \
    \mathcal{R}_{\mu\nu\sigma}{}^\lambda \, \Sigma^{\mu\nu}{}_{\rho\lambda} \ - \
    \frac{1}{k} \, g_{\rho\sigma} \, \mathcal{R}_{\mu\nu\alpha\beta} \, \Sigma^{\mu\nu\alpha\beta} \ = \ 0,
\label{Einsteineqn-LL}\\ [.2cm]    
&& \nabla_\mu \Sigma^{\mu\nu\alpha}{}_{\beta}
\ - \ \half \, Q_{\mu\lambda}{}^\lambda \, \Sigma^{\mu\nu\alpha}{}_{\beta}
\ + \ T_{\sigma\mu}{}^\sigma \, \Sigma^{\mu\nu\alpha}{}_{\beta}
\ - \ \half \, T_{\mu\sigma}{}^\nu \, \Sigma^{\mu\sigma\alpha}{}_{\beta} \ = \ 0,
\label{EOMGamma-LL}  
\eea
with $\nabla_\mu$ the covariant derivative, $Q_\mnr = -\nabla_\mu g_{\nu\rho}$ the non-metricity tensor 
and $T_{\mu\nu}{}^\rho = 2\Gamma_{[\mu\nu]}{}^\rho$
the torsion of the general connection $\Gamma_{\mu\nu}{}^\rho$.

Both equations (\ref{Einsteineqn-LL}) and (\ref{EOMGamma-LL}) can be simplified considerably.
Taking the trace
of (\ref{Einsteineqn-LL}) tells us that
$\mathcal{R}_{\mu\nu\alpha\beta} \, \Sigma^{\mu\nu\alpha\beta} =0$ in any dimension
except $D=2k$, such that the traceless part of the metric equation in $D\neq 2k$ is given by
\be
\mathcal{R}_{\mu\nu\rho}{}^\lambda \, \Sigma^{\mu\nu}{}_{\sigma\lambda} \ + \
    \mathcal{R}_{\mu\nu\sigma}{}^\lambda \, \Sigma^{\mu\nu}{}_{\rho\lambda} \ = \ 0.
\ee
On the other hand, splitting the general connection  $\Gamma_{\mu\nu}{}^\rho$ in its
Levi-Civita part and a tensorial part,
\be
\Gamma_{\mu\nu}{}^\rho \ =  \ \mathring\Gamma_{\mu\nu}{}^\rho \ + \ K_{\mu\nu}{}^\rho,
\label{Gammadecomp}
\ee
the connection equation (\ref{EOMGamma-LL}) can be written in the simple form 
\be
\mathring\nabla_\mu \Sigma^{\mu\nu\alpha}{}_{\beta}
\ + \ K_{\mu\rho}{}^\alpha \, \Sigma^{\mu\nu\rho}{}_{\beta} 
\ - \ K_{\mu\beta}{}^\rho \, \Sigma^{\mu\nu\alpha}{}_{\rho} \ = \ 0,
\label{EOMGamma-LL2}
\ee
where $\mathring\nabla$ is the covariant derivative with respect to the Levi-Civita connection.
It is worth observing that both (\ref{EOMGamma-LL}) and (\ref{EOMGamma-LL2}), written in terms
of $\Sigma^{\mu\nu\alpha}{}_{\beta}$, are in fact completely general, for any lagrangian of the type
$\cL(g_\mn, \cR_\mnr{}^\lambda)$, not just for the Lovelock lagrangian (\ref{Lovelock-Palatini}).

Furthermore, using the antisymmetry of the Lovelock $\Sigma^{\mu\nu}{}_{\alpha\beta}$
in the lower indices, 
it is easy to show that from (\ref{EOMGamma-LL2}) one can deduce the necessary (thought not
sufficient) condition for the connection,
\be
\Bigl(K_{\mu\rho\alpha} \ + \ K_{\mu\alpha\rho}\Bigr) \, \Sigma^{\mu\nu\rho}{}_\beta \ + \
\Bigl(K_{\mu\rho\beta} \ + \ K_{\mu\beta\rho}\Bigr) \, \Sigma^{\mu\nu\rho}{}_\alpha \ = \ 0.
\label{necessary}
\ee

\sect{Levi-Civita as a solution and projective symmetry}
\label{lc}

It is well known \cite{ESJ, BJB, DP} that the Levi-Civita connection is a solution
of all order Lovelock terms in arbitrary dimensions ($D\geq2k$). The proof is particularly easy
in terms of $\Sigma^{\mu\nu}{}_{\alpha\beta}$ and the decomposition (\ref{Gammadecomp}): since for
the Levi-Civita connection we have that $\mathring K_{\mu\nu}{}^\rho\equiv 0$ identically,
the connection  equation (\ref{EOMGamma-LL2}) takes the form
\be
0 \ = \ \mathring\nabla_\mu \mathring\Sigma^{\mu\nu\alpha}{}_{\beta}
\ = \ k(k-1) \
\delta^{\mu\nu\rho_2\lambda_2 \dots \rho_k\lambda_k}_{\alpha\beta\gamma_2\epsilon_2 \dots \gamma_k\epsilon_k} \
   \mathring\nabla_\mu\mathring{\mathcal{R}}_{\rho_2\lambda_2}{}^{\gamma_2\epsilon_2} \,
   \mathring{\mathcal{R}}_{\rho_3\lambda_3}{}^{\gamma_3\epsilon_3}\dots \,
   \mathring{\mathcal{R}}_{\rho_k\lambda_k}{}^{\gamma_k\epsilon_k},
\ee
which is automatically satisfied due to the second Bianchi identity for the Levi-Civita 
Riemann tensor, $\mathring\nabla_{[\mu} \mathring{\cal{R}}_{\nu\rho]\lambda}{}^\sigma = 0$.
On the other hand, the metric equation (\ref{Einsteineqn-LL}),
\be
\mathring{\mathcal{R}}_{\mu\nu\rho}{}^\lambda \, \mathring\Sigma^{\mu\nu}{}_{\sigma\lambda} \ + \
    \mathring{\mathcal{R}}_{\mu\nu\sigma}{}^\lambda \, \mathring\Sigma^{\mu\nu}{}_{\rho\lambda} \ - \
    \frac{1}{k} \, g_{\rho\sigma} \, \mathring{\mathcal{R}}_{\mu\nu\alpha\beta}
    \, \mathring\Sigma^{\mu\nu\alpha\beta} \ = \ 0,
\ee
reduces to the equation of motion for $g_{\mu\nu}$ of the Lovelock action in the metric formalism,
without imposing any extra conditions on the connection. This proves that the Levi-Civita
connection is a consistent truncation in metric-affine Lovelock theory \cite{DP}. 

It is straightforward to see \cite{JJOS} that the Lovelock action (\ref{Lovelock-Palatini})
is also invariant under projective transformations,
\be \Gamma_\mn{}^\rho \ \rightarrow \ \bar\Gamma_\mn{}^\rho
\ = \ \Gamma_\mn{}^\rho \ + \ A_\mu \,\delta_\nu^\rho,
\label{projtransf2} 
\ee
in fact almost trivially so. Indeed, since the Riemann tensor transforms under projective
transformations as
\be
 \cR_ \mnr{}^\lambda (\Gamma)  \ \rightarrow  \ \bar\cR_ \mnr{}^\lambda  (\bar\Gamma)
   \ = \ \cR_ \mnr{}^\lambda  (\Gamma)\, +  \, F_\mn(A)  \, \delta_\rho^\lambda,
\ee
with  $F_\mn(A) = 2\partial_{[\mu} A_{\nu]}$, the Lovelock $\Sigma$-tensor (\ref{LL-Sigma})
is invariant under (\ref{projtransf2}),
\bea
\Sigma^{\mu\nu}{}_{\alpha\beta} \ \rightarrow \ 
\bar \Sigma^{\mu\nu}{}_{\alpha\beta} 
 &=& k \ \delta^{\mu\nu\rho_2\lambda_2 \dots \rho_k\lambda_k}_{\alpha\beta\gamma_2\epsilon_2 \dots \gamma_k\epsilon_k} \
    \Bigl[\mathcal{R}_{\rho_2\lambda_2}{}^{\gamma_2\epsilon_2}
           + F_{\rho_2\lambda_2} g^{\gamma_2\epsilon_2} \Bigr] \dots 
 \Bigl[\mathcal{R}_{\rho_k\lambda_k}{}^{\gamma_k\epsilon_k}
 + F_{\rho_k\lambda_k} g^{\gamma_k\epsilon_k} \Bigr] \nnw
    &=& \Sigma^{\mu\nu}{}_{\alpha\beta},
\eea
due to the antisymmetry of the $\delta$-tensor and the symmetry of the metric.
For the same reason we have that
\be
\bar{\mathcal{R}}_{\mu\nu}{}^{\alpha\beta} \, \bar\Sigma^{\mu\nu}{}_{\alpha\beta}
\ = \ \Bigl[\mathcal{R}_{\mu\nu}{}^{\alpha\beta}
    \,+  \, F_\mn(A)  \, g^{\alpha\beta}  \Bigr]\, \Sigma^{\mu\nu}{}_{\alpha\beta} 
\ = \ \mathcal{R}_{\mu\nu}{}^{\alpha\beta} \, \Sigma^{\mu\nu}{}_{\alpha\beta} 
\ee
and hence the action (\ref{Lovelock-Palatini2}) is invariant.

Just as in the case of the Einstein-Hilbert action, the projective invariance of the Lovelock
action allows for solutions of the connection equation of the type (\ref{PalatiniGamma}).
Moreover, given an affine connection $\Gamma_{\mu\nu}{}^\rho$ (not necessarily Levi-Civita) that
is a solution to the equations (\ref{Einsteineqn-LL}) and (\ref{EOMGamma-LL2}), we can always
build a new connection
\be
  \bGamma_\mn{}^\rho \ = \ {\Gamma}_\mn{}^\rho \ + \ A_\mu \,\delta_\nu^\rho,
  \label{Palatiniadditivity}
\ee
that also solves the same equations of motion. 
Just as for the Einstein-Hilbert case, the projective symmetry of the action guarantees
that $\bar\Gamma_\mn{}^\rho$ and  $\Gamma_\mn{}^\rho$ are physically indistinguishable.
Therefore, the space of affine connections allowed by the equations of motion of Lovelock
theories consists of a set of equivalence classes $[\Gamma]$, where different elements 
within the same class are related as in (\ref{Palatiniadditivity}) with arbitrary $A_\mu$,
while connections from  different classes describe different physics.

It is sometimes said that the metric and the Palatini formalism are equivalent for all Lovelock
theories. However strictly speaking this would only be the case if the space of solutions
contains $[\mathring\Gamma]$ as unique equivalence class. In other words, if the only allowed
solutions are of
the form (\ref{PalatiniGamma}). While this is clearly the case for the Einstein-Hilbert action
\cite{Pons, BJJOSS, JJOS}, it remains an open question for the higher-order Lovelock theories.
What distinguishes the Einstein-Hilbert action form the rest of the Lovelock terms is the fact
that $\Sigma^{\mu\nu}{}_{\alpha\beta}$ does not depend on $\Gamma$ and therefore the connection
equation (\ref{EOMGamma-LL2}) is an algebraic equation. In the general case, however,
(\ref{EOMGamma-LL2}) is a non-linear second-order differential equation for $\Gamma$.

We will show that in general the metric and the Palatini formalisms are not equivalent
for higher-order Lovelock theories, by presenting a concrete counterexample for a specific
theory: the Weyl connection for the four-dimensional Gauss-Bonnet term. 

\sect{The Weyl connection as a solution}
\label{Wsolution}

We now consider the $D$-dimensional second-order Lovelock term, also known as the
Gauss-Bonnet action,\footnote{Written out explicitly in terms of the curvature  tensors,
the action (\ref{Gauss-Bonnet-Palatini}) is given by
\[
\mathcal{L} \ = \ \frac{1}{3!} \, \sqrtg \ \Bigl[\mathcal{R}^2
\ - \ \mathcal{R}^{(1)}_{\mu\nu} \mathcal{R}^{(1)\nu\mu}
\ + 2 \ \mathcal{R}^{(1)}_{\mu\nu} \mathcal{R}^{(2)\nu\mu}
\ - \ \mathcal{R}^{(2)}_{\mu\nu} \mathcal{R}^{(2)\nu\mu}
\ + \ \mathcal{R}_{\mu\nu\rho\lambda}\mathcal{R}^{\rho\lambda\mu\nu}\Bigr],
\nonumber
\]
where $\mathcal{R}^{(1)}_{\mu\nu} = \mathcal{R}_{\mu\lambda\nu}{}^\lambda$ is the Ricci tensor,
$\mathcal{R}^{(2)}_{\mu\nu} = g^{\rho\lambda} \mathcal{R}_{\mu\rho\lambda\nu}$ the co-Ricci tensor and
$\mathcal{R} = g^{\mu\nu} \mathcal{R}^{(1)}_{\mu\nu}$ the Ricci scalar. However, we will prefer to
work throughout this paper with the $\Sigma$-tensor notation.}
\bea
\mathcal{L}^{(D)}_{\mathrm{GB}} (g, \Gamma)
\ = \ \sqrtg \ \delta_{\alpha\beta\gamma\epsilon}^{\mu\nu\rho\lambda} \, \,
\mathcal{R}_{\mu\nu}{}^{\alpha\beta}(\Gamma) \, \mathcal{R}_{\rho\lambda}{}^{\gamma\epsilon}(\Gamma),
\label{Gauss-Bonnet-Palatini}
\eea
such that the $\Sigma$-tensor (\ref{LL-Sigma}) takes the form
\be
\Sigma^{\mu\nu}{}_{\alpha\beta} \ =
\  2\, \delta^{\mu\nu\rho\lambda}_{\alpha\beta\gamma\epsilon} \, \mathcal{R}_{\rho\lambda}{}^{\gamma\epsilon}(\Gamma).
\ee
We will try to find a non-trivial connection $\Gamma_{\mu\nu}{}^\rho$ ({\it i.e.}\ not of the
form (\ref{PalatiniGamma})) that solves the metric and connection equations
(\ref{Einsteineqn-LL}) and (\ref{EOMGamma-LL2}) for the case $k=2$. 

Our Ansatz will be the generalised Weyl connection,
\be
\tilde \Gamma_{\mu\nu}{}^\rho \ = \ \mathring\Gamma_{\mu\nu}{}^\rho
\ + \ A_\mu \, \delta_\nu^\rho \ + \ B_\nu \, \delta_\mu^\rho
  \ - \ {C}^\rho g_\mn,
\label{generalisedWeyl}
\ee
characterised by three arbitrary vector fields ${A}_\mu$, ${B}_\mu$ and 
${C}_\mu$. Strictly speaking, $A_\mu$ represents the projective symmetry of the action and can 
be gauged away completely. However for future reference, we prefer to maintain the calculation 
general for the moment. The Riemann and the $\Sigma$-tensor for this connection are then given
by
\bea
\tilde{\mathcal{R}}_\mnr{}^\lambda &=& \mathring{\mathcal{R}}_\mnr{}^\lambda
\ + \ F_\mn(A) \, \delta_\rho^\lambda
\ + \ \Bigl[\mathring\nabla_\mu B_\rho -  B_\mu B_\rho\Bigr] \delta^\lambda_\nu 
\ - \ \Bigl[\mathring\nabla_\nu B_\rho -  B_\nu B_\rho\Bigr] \delta^\lambda_\mu \nnw
&& \ - \  \Bigl[\mathring\nabla_\mu C^\lambda -  C_\mu C^\lambda\Bigr] g_{\nu\rho}
\ + \  \Bigl[\mathring\nabla_\nu C^\lambda -  C_\nu C^\lambda\Bigr] g_{\mu\rho}
\ - \ B_\sigma C^\sigma \Bigl[ \delta_\mu^\lambda g_{\nu\rho} \, - \,  \delta_\nu^\lambda g_{\mu\rho} \Bigr]
\nnw
\tilde{\Sigma}^\mn{}_{\alpha\beta} &=& \mathring{\Sigma}^\mn{}_{\alpha\beta} 
     \ + \ \half (D-3) \delta^{\mu\nu\rho}_{\alpha\beta\gamma} 
        \Bigl[\mathring\nabla_\rho B^\gamma -  B_\rho B^\gamma \Bigr] \nnw
 &&   \ + \ \half (D-3) \delta^{\mu\nu\rho}_{\alpha\beta\gamma} 
        \Bigl[\mathring\nabla_\rho C^\gamma -  C_\rho C^\gamma\Bigr]
    \ + \ \tfrac{1}{6} (D-2)(D-3) \delta^\mn_{\alpha\beta} B_\sigma C^\sigma.
\eea
Plugging the Ansatz (\ref{generalisedWeyl}) in the necessary condition (\ref{necessary}),
we find that
\bea
0 &\equiv& \Bigl(\tilde K_{\mu\rho\alpha}
\ + \ \tilde K_{\mu\alpha\rho}\Bigr) \, \tilde \Sigma^{\mu\nu\rho}{}_\beta \ + \
\Bigl(\tilde K_{\mu\rho\beta} \ + \ \tilde K_{\mu\beta\rho}\Bigr) \, \tilde \Sigma^{\mu\nu\rho}{}_\alpha
\nnw
 &=& 2\Bigl(B_\rho - C_\rho\Bigr) \tilde\Sigma_{(\alpha}{}^{\nu\rho}{}_{\beta)}
   \ + \  \Bigl(B_{(\alpha} - C_{(\alpha}\Bigr) \tilde\Sigma^{\mn}{}_{\beta)\mu},
\eea
which is satisfied only when $B_\mu = C_\mu$. If we then gauge fix the projective symmetry
by choosing also $A_\mu = B_\mu$,  so that we can write the Ansatz (\ref{generalisedWeyl})
as a non-integrable Weyl connection,
\be
\tilde \Gamma_{\mu\nu}{}^\rho \ = \ \mathring\Gamma_{\mu\nu}{}^\rho
\ + \ B_\mu \, \delta_\nu^\rho \ + \ B_\nu \, \delta_\mu^\rho
  \ - \ B^\rho g_\mn,
\label{Weyl}
\ee
with $B_\mu$ for the moment an arbitrary vector field, whose precise form should be determined
by the equations of motion. Filling in the Ansatz (\ref{Weyl}) into the
connection equation (\ref{EOMGamma-LL2}) yields
\bea
0 &\equiv&  \mathring\nabla_\mu \tilde \Sigma^{\mu\nu}{}_{\alpha\beta}
\ + \ \tilde K_{\mu\rho\alpha} \, \tilde \Sigma^{\mu\nu}{}_{\lambda\beta}\, g^{\rho\lambda} 
\ - \ \tilde K_{\mu\beta\rho} \, \tilde \Sigma^{\mu\nu}{}_{\alpha\lambda} \, g^{\rho\lambda}  \nnw
&=&
 \tfrac{1}{12} (D-4) \Bigl[ \,2B_{[\beta} \mathring{\mathcal{R}} \, \delta^\nu_{\alpha]}
 \ + \ 4 B^\lambda  \mathring{\mathcal{R}}_{\lambda[\alpha} \delta^\nu_{\beta]}
 \ - \ 2 B^\lambda \mathring{\mathcal{R}}_{\alpha\beta\lambda}{}^\nu\Bigr] \nnw
 && 
 \ + \ \tfrac{1}{6} (D-4)(D-3) \Bigl[ \,2 B_\rho \mathring\nabla_{[\alpha} B^\rho \delta^\nu_{\beta]}
  \ - \ 2 B_{[\alpha} \mathring\nabla_{|\rho|} B^\rho \delta^\nu_{\beta]}
  \ + \ 2 B_{[\alpha} \mathring\nabla_{\beta]} B^\nu \Bigr] \nnw
  && 
  \ - \ \tfrac{1}{6} (D-4)(D-3)(D-2) \, B_\sigma B^\sigma B_{[\alpha} \delta^\nu_{\beta]},
\eea
which is satisfied for arbitrary vector fields $B_\mu$ in $D=4$. On the other hand,
the metric equation (\ref{Einsteineqn-LL})  becomes
\bea
0 &=& \tilde{\mathcal{R}}_{\mu\nu\alpha}{}^\lambda \, \tilde \Sigma^{\mu\nu}{}_{\beta\lambda} \ + \
   \tilde{ \mathcal{R}}_{\mu\nu\beta}{}^\lambda \, \tilde \Sigma^{\mu\nu}{}_{\alpha\lambda} \ - \
    \tfrac{1}{2} \, g_{\alpha\beta} \, \tilde{\mathcal{R}}_{\mu\nu\rho\lambda} \,
    \tilde \Sigma^{\mu\nu\rho\lambda} \nnw
&=&  \mathring{\mathcal{R}}_{\mu\nu\alpha}{}^\lambda \, \mathring \Sigma^{\mu\nu}{}_{\beta\lambda} \ + \
   \mathring{ \mathcal{R}}_{\mu\nu\beta}{}^\lambda \, \mathring \Sigma^{\mu\nu}{}_{\alpha\lambda} \ - \
    \tfrac{1}{2} \, g_{\alpha\beta} \, \mathring{\mathcal{R}}_{\mu\nu\rho\lambda} \,
    \mathring \Sigma^{\mu\nu\rho\lambda}\nnw
&& \ + \ \tfrac{1}{3} (D-4) \Bigl[ \mathring\nabla_{(\alpha} B_{\beta)} \mathring{\mathcal{R}} 
\ + \ 2\mathring\nabla_\mu B^\mu ( \mathring{\mathcal{R}}_{\alpha\beta}
 - \half g_{\alpha\beta}\mathring{\mathcal{R}})
\ + \ 2\mathring\nabla^\mu B^\nu \mathring{\mathcal{R}}_{\mu(\alpha\beta)\nu} \nnw
&&\hsp{2cm} \ - \ 2 \mathring\nabla_{(\alpha} B^\mu \mathring{\mathcal{R}}_{\beta)\mu}
\ - \ 2 \mathring\nabla_{\mu} B_{(\alpha} \mathring{\mathcal{R}}_{\beta)}{}^{\mu}
\ + \ 2 \mathring\nabla_{\mu} B_\nu \mathring{\mathcal{R}}^\mn g_{\alpha\beta} \Bigr]\nnw
&& \ + \ \tfrac{1}{3} (D-4) \Bigl[
(D-5) B_\mu B^\mu (\mathring{\mathcal{R}}_{\alpha\beta}  - \half g_{\alpha\beta}\mathring{\mathcal{R}})
\ - \ B_\alpha B_\beta \mathring{\mathcal{R}}
\ - \ 2 B^\mu B^\nu \mathring{\mathcal{R}}_{\mu\nu}g_{\alpha\beta} \nnw
&&\hsp{2cm} \ + \ 4 (D-3)B^\mu B_{(\alpha} \mathring{\mathcal{R}}_{\beta)\mu}
\ - \ 2  B^\mu B^\nu \mathring{\mathcal{R}}_{\mu(\alpha\beta)\nu} \Bigr] \nnw
&& \ + \ \tfrac{1}{3} (D-4)(D-3) \Bigl[
\, 2\mathring\nabla_{(\alpha} B_{\beta)} \mathring\nabla_{\mu} B^{\mu}
\ - \ 2 \mathring\nabla_\mu B_{(\alpha} \mathring\nabla_{\beta)} B^{\mu} \nnw
&&\hsp{2cm}
\ - \  \mathring\nabla_{\mu} B^{\mu}\mathring\nabla_{\nu} B^{\nu}g_{\alpha\beta}
\ + \ \mathring\nabla_{\mu} B^{\nu}\mathring\nabla_{\nu} B^{\mu}g_{\alpha\beta}\Bigr]\nnw
&& \ + \ \tfrac{1}{3} (D-4)(D-3) \Bigl[
(D-4) \mathring\nabla_{(\alpha} B_{\beta)}B_\mu B^\mu
\ - \ 2 \mathring\nabla_{\mu} B^\mu B_\alpha B_\beta
\ + \ 2  \mathring\nabla_{\mu} B_{(\alpha} B_{\beta)} B^\mu \nnw
&& \hsp{2cm}
\ + \ 2 B^\mu B_{(\alpha}  \mathring\nabla_{\beta)} B_\mu
\ - \ 2 B^\mu B^\nu \mathring\nabla_{\mu} B_\nu g_{\alpha\beta}
\ + \ (D-4) B_\mu B^\mu  \mathring\nabla_{\nu} B^\nu g_{\alpha\beta}\Bigr]\nnw
&&  + \ \tfrac{1}{12} (D-4)(D-3)(D-2) \Bigl[ 4B_\mu B^\mu B_\alpha B_\beta
\ + \ (D-5) B_\mu B^\mu B_\nu B^\nu  g_{\alpha\beta}\Bigr],
\eea
which in $D=4$ reduces to equation of motion for $g_\mn$
in the metric formalism,
\be
\mathring{\mathcal{R}}_{\mu\nu\alpha}{}^\lambda \, \mathring \Sigma^{\mu\nu}{}_{\beta\lambda} \ + \
   \mathring{ \mathcal{R}}_{\mu\nu\beta}{}^\lambda \, \mathring \Sigma^{\mu\nu}{}_{\alpha\lambda} \ - \
    \tfrac{1}{2} \, g_{\alpha\beta} \, \mathring{\mathcal{R}}_{\mu\nu\rho\lambda} \,
        \mathring \Sigma^{\mu\nu\rho\lambda}
\ = \ 0.
\ee
In other words, the Weyl connection (\ref{Weyl}) is a solution of four-dimensional
metric-affine Gauss-Bonnet gravity for any $g_\mn$ that satisfies the equations
of the metric formalism.


\noindent
\sect{A vector symmetry of $D=4$ Gauss-Bonnet theory}
\label{symmetry}

In section \ref{lc} we have seen that the existence of the nontrivial connection
(\ref{PalatiniGamma}) $\bGamma_\mn{}^\rho =  {\mathring\Gamma}_\mn{}^\rho
+A_\mu \delta_\nu^\rho$, as a solution in any metric-affine Lovelock theory is a
consequence of the projective symmetry $\Gamma_\mn{}^\rho \rightarrow
\bar\Gamma_\mn{}^\rho = \Gamma_\mn{}^\rho + A_\mu \delta_\nu^\rho$. In this section
we will argue that our new solution (\ref{Weyl}) is also related to a symmetry,
namely the conformal invariance of the four-dimensional Gauss-Bonnet action.

Conformal invariance and Weyl transformation have not been studied much in the context
of metric-affine gravity. In \cite{CDV2} conformal rescalings of the metric are used to
discuss the relations between the metric and Palatini formalism of in $f(R)$ gravity
in both the Einstein and the Jordan frame. More recently, in \cite{IK} a detailed
classification was given of the metric-affine theories in terms of their scale
invariance under rescalings of the metric, the coframe and/or the connection.

It is well known that the metric Gauss-Bonnet theory in $D=4$ is invariant under
conformal transformations of the metric,
\be
g_\mn \ \rightarrow \ \tilde g_\mn \ = \ e^{2\phi} \ g_\mn,
\label{conftransfg}
\ee
which on its turn change the Christoffel symbols as
\be
\mathring \Gamma_\mn{}^\rho \ \rightarrow \ \tilde \Gamma_\mn{}^\rho
\ = \ \mathring \Gamma_\mn{}^\rho \ + \ \partial_\mu \phi \, \delta_\nu^\rho \
+ \ \partial_\nu \phi \, \delta_\mu^\rho
\ - \ \partial^\rho \phi \, g_\mn.
\label{conftransfGamma}
\ee
On the other hand, as any metric-affine quadratic curvature term \cite{BFFV}, the
four-dimensional metric-affine Gauss-Bonnet theory is easily seen to have conformal
weight zero , {\it i.e.}\ to be invariant under the conformal transformations
(\ref{conftransfg}) of the metric, though in this context without a accompanying
transformation in the affine connection, as the latter is independent of the metric.

The invariance of the $D=4$ metric-affine Gauss-Bonnet term under the metric transformation
(\ref{conftransfg}) shows that in the metric formalism the transformation of the metric
(\ref{conftransfg}) and of the connection (\ref{conftransfGamma}) are in fact quite
independent of each other: (\ref{conftransfg}) acts effectively only on the explicit metrics
in the contraction of the Riemann tensors and  the effect of (\ref{conftransfGamma}) remains
constrained to the curvature tensors. One could therefore ask the question whether the
metric-affine Gauss-Bonnet action is also invariant under (something similar to) the
transformation (\ref{conftransfGamma}), independently of a metric transformation. 

In \cite{BK, BK2, BHK} it was already observed that actions with Gauss-Bonnet-like
quadratic curvature invariants ({\it i.e.}\ general combinations of quadratic contractions of
the Riemann tensor, that reduce to the metric Gauss-Bonnet action when the Levi-Civita
connection is imposed), when equipped with the (non-integrable) Weyl connection (\ref{Weyl}),
can be written as the standard (Levi-Civita) Gauss-Bonnet action plus a series of non-minimal
coupling terms for the Weyl field $B_\mu$, plus a kinetic term $F_\mn(B) F^\mn(B)$. Curiously
enough, the non-minimal couplings vanish precisely in $D=4$ and the kinetic term is multiplied
by a coefficient that vanishes when the parameters of the extended Gauss-Bonnet term are
chosen such that the action is the actual metric-affine Gauss-Bonnet term
(\ref{Gauss-Bonnet-Palatini}). In other words, the metric-affine Gauss-Bonnet action
(\ref{Gauss-Bonnet-Palatini}) does not see the difference between the substituting the
Weyl or the Levi-Civita connection. 

Inspired by this and by the fact that in the previous
section we found that the integrable Weyl connection (\ref{Weyl}) is a solution to the
metric and the connection equation, it seems logical to check the invariance of the
action (\ref{Gauss-Bonnet-Palatini}) under the transformation 
\be
\Gamma_\mn{}^\rho \ \rightarrow \ \tilde \Gamma_\mn{}^\rho
\ = \ \Gamma_\mn{}^\rho \ + \ B_\mu  \, \delta_\nu^\rho \ + \ B_\nu  \, \delta_\mu^\rho
\ - \ B^\rho \, g_\mn,
\label{vectortransfGamma}
\ee
not just as a deformation of the Levi-Civita connection (as in \cite{BK, BK2, BHK}), but
as a transformation acting on general connections in the action (\ref{Gauss-Bonnet-Palatini}),
much in the same way as the projective transformations (\ref{projtransf}). Note that the
$B_\mu  \delta_\nu^\rho$ term can be undone by a projective transformation with parameter
$-B_\mu$, so we can actually simplify the transformation
(\ref{vectortransfGamma}) to
\be
\Gamma_\mn{}^\rho \ \rightarrow \ \tilde \Gamma_\mn{}^\rho
\ = \ \Gamma_\mn{}^\rho \ + \ B_\nu  \, \delta_\mu^\rho \ - \ B^\rho \, g_\mn.
\label{vectortransfGamma2}
\ee
Up to boundary terms coming from integrating by parts, the four-dimensional action then
transforms as
\be
\mathcal{L}_{\mathrm{GB}} (g, \Gamma)\ \rightarrow \
\tilde{\mathcal{L}}_{\mathrm{GB}} (g, \tilde \Gamma),
\ee
with
\bea
\tilde{\mathcal{L}}_{\mathrm{GB}} (g, \tilde \Gamma)
&=&  \mathcal{L}_{\mathrm{GB}} (g, \Gamma)
\ - \ 4 B^\mu B^\nu \Bigr[ \mathcal{R}^{(1)}_{\mu\nu} \, + \, \mathcal{R}^{(2)}_{\mu\nu}\Bigr]
\nnw
&&- \ 2Q^\mnr\Bigr[ B_\mu (\mathcal{R}^{(1)}_{\nu\rho} \, + \, \mathcal{R}^{(2)}_{\nu\rho})
\ + \  B^\lambda (\mathcal{R}_{\lambda\nu\mu\rho}\, + \, \mathcal{R}_{\lambda\nu\rho\mu})
 \nnw
&& \hsp{1.5cm}
\ - \  B^\lambda B_\nu ( Q_{\lambda\mu\rho}\, - \, 2 Q_{\rho\lambda\mu})
\ - \ B_\mu B_\nu (Q^{(1)}_\rho \, - \,  Q^{(2)}_\rho)
\nnw
&& \hsp{1.5cm}
\ + \ 2B_\mu \nabla_\nu B_\rho   \ + \ 4B_\nu \nabla_\rho B_\mu 
\ + \ 2 B_\nu B^\lambda T_{\lambda\rho\mu}
\ + \ 2 B_\mu B_\nu T_{\rho\lambda}{}^\lambda \Bigr] \\ [.2cm]
&& - \ 2Q^{(1)\mu} \Bigl[
 B^\nu (\mathcal{R}^{(1)}_{\nu\mu} \, - \, \mathcal{R}^{(2)}_{\nu\mu}
           \, - \,  g_{\nu\mu} \mathcal{R})
\ - \ 2 B_\mu B_\nu B^\nu
\nnw
&& \hsp{1.5cm}
\ - \ 2 B_\mu \nabla_\nu B^\nu
\ + \ 2 B^\nu \nabla_\nu B_\mu
\ + \ 3 B_\mu B^\nu Q^{(2)}_\nu\Bigr] \nnw
&& +\ 2 Q^{(2)\mu} \Bigl[
B^\nu (\mathcal{R}^{(1)}_{\nu\mu} \, + \, \mathcal{R}^{(2)}_{\nu\mu})
\ + \ 2 B_\mu \nabla_\nu B^\nu
\ + \ 2 B^\nu \nabla_\nu B_\mu
\ + \ 2 B_\mu B^\nu T_{\nu\lambda}{}^\lambda   \Bigr], \nonumber
\eea
where $\mathcal{R}^{(1)}_{\mu\nu} = \mathcal{R}_{\mu\lambda\nu}{}^\lambda$ is the Ricci tensor,
$\mathcal{R}^{(2)}_{\mu\nu} = g^{\rho\lambda} \mathcal{R}_{\mu\rho\lambda\nu}$ the co-Ricci tensor,
$\mathcal{R} = g^{\mu\nu} \mathcal{R}_{\mu\nu}$ the Ricci scalar and
$Q^{(1)}_\mu = Q_{\mu\lambda}{}^\lambda$ and $Q^{(2)}_\mu = Q^\lambda{}_{\lambda\mu}$
the two traces of the non-metricity tensor $Q_\mnr = - \nabla_\mu g_{\nu\rho}$.

We can see then that in fact the four-dimensional metric-affine Gauss-Bonnet term
(\ref{Gauss-Bonnet-Palatini}) with general connection $\Gamma_\mn{}^\rho$ is not invariant
under (\ref{vectortransfGamma2}). However, taking into account that the Ricci and the
co-Ricci tensor are in general related to each other as
\be
\mathcal{R}^{(2)}_{\mu\nu} \ = \ - \mathcal{R}^{(1)}_{\mu\nu}
\ + \ g^{\rho\lambda} \nabla_\mu Q_{\rho\nu\lambda}
\ + \ g^{\rho\lambda} \nabla_\rho Q_{\mu\nu\lambda}
\ + \ g^{\rho\lambda} T_{\mu\rho}{}^\sigma Q_{\sigma\nu\lambda},
\ee
it is clear that the difference between $\mathcal{L}_{\mathrm{GB}} (g, \Gamma)$ and
$\tilde{\mathcal{L}}_{\mathrm{GB}} (g, \tilde \Gamma)$ is proportional to the non-metricity
tensor,
its derivatives and its traces. In other words, the transformation (\ref{vectortransfGamma2})
is indeed a symmetry, not of the full four-dimensional metric-affine Gauss-Bonnet action,
but of the restriction of this theory to the
subset of metric-compatible connections, which turns out to be a consistent
truncation of the full theory (see Appendix \ref{Q=0}).
The symmetry transformation (\ref{vectortransfGamma2})
not only generalises the results of \cite{BK, BK2, BHK}, but also explains why
the Weyl connection (\ref{Weyl}) appears as a solution to the Palatini formalism in
the four-dimensional Gauss-Bonnet action: it arises by acting on the Levi-Civita solution
first with the new vector symmetry (\ref{vectortransfGamma2}) and then with a projective
transformation (\ref{Palatiniadditivity}) with the same parameter. Note that the order of
these transformations is important, as the vector transformation is only a symmetry on the
subset of metric-compatible connections. This subset itself is not
invariant under projective transformations, since any projective transformation
necessarily induces a non-trivial non-metricity:
$Q_\mnr \rightarrow \bar Q_\mnr = Q_\mnr + 2 A_\mu g_{\nu\rho}$.

\noindent
\sect{Conclusions}
\label{conclusions}

While looking for solutions of the connection equation of metric-affine Gauss-Bonnet theory
$\mathcal{L}_{\mathrm{GB}}(g, \Gamma)$ (\ref{Gauss-Bonnet-Palatini}), we have identified a number of
transformations in the theory. Besides the invariance under projective transformations,
\be
\Gamma_\mn{}^\rho \ \rightarrow \ \bar \Gamma_\mn{}^\rho
\ = \ \Gamma_\mn{}^\rho \ + \ A_\mu\, \delta_\nu^\rho,
\label{proj2}
\ee
present in any dimension, we also found a vector transformation
\be
\Gamma_\mn{}^\rho \ \rightarrow \ \hat \Gamma_\mn{}^\rho
\ = \ \Gamma_\mn{}^\rho\ + \ B_\nu\, \delta_\mu^\rho
\ - \ B^\rho g_\mn,
\label{weyl2}
\ee
which is a symmetry specifically in four-dimensions and only if we consider the theory to
be restricted to metric-compatible connections ($\mathcal{L}_{\mathrm{GB}}|_{Q=0}$).
However, this vector
transformation will play an important role in  the full (four-dimensional) theory
$\mathcal{L}_{\mathrm{GB}}$.

To our knowledge, this vector symmetry (\ref{weyl2}) of the  truncated theory
$\mathcal{L}_{\mathrm{GB}}|_{Q=0}$ is new, although a special case was already observed in
\cite{BK, BK2, BHK}. Both the $A_\mu$ and $B_\mu$ transformations seem somehow to
be related to the conformal invariance of the four-dimensional Gauss-Bonnet action
in the metric formalism,
\be
g_\mn \ \rightarrow \ \tilde g_\mn \ = \ e^{2\phi} \, g_\mn,
\hsp{1cm}
\mathring \Gamma_\mn{}^\rho \ \rightarrow \ \tilde \Gamma_\mn{}^\rho
\ = \ \mathring \Gamma_\mn{}^\rho \ + \ \partial_\mu \phi \, \delta_\nu^\rho \
+ \ \partial_\nu \phi \, \delta_\mu^\rho
\ - \ \partial^\rho \phi \, g_\mn.
\label{conftransf}
\ee
Note that the conformal weight of the four-dimensional Gauss-Bonnet term is zero, both
in the metric as in the metric-affine formalism. Therefore, in the metric case,
the $(\partial\phi)$-terms that come from the transformation of
the Levi-Civita connection cancel out amongst each other, and hence the transformation rules
(\ref{conftransf}) for the metric and the connection do not interfere with each other in
the variation of 
the action (\ref{Gauss-Bonnet-Palatini}). Moreover, in the metric-affine formalism, where
the metric and the affine connection are independent variables, one can separate
both transformations completely, finding that the action is invariant under both of them
separately, at least in the subset of metric-compatible connections. 
The remarkable thing is that the metric-compatible Gauss-Bonnet term allows
not only for integrable Weyl vectors $B_\mu = \partial_\mu \phi$, but also for non-integrable
ones, $B_\mu \neq \partial_\mu \phi$, as the transformation is no longer related to a conformal
transformation
of the metric.

To understand the mathematical structure of the space of solutions of the full 
four-dimensional Gauss-Bonnet action $\mathcal{L}_{\mathrm{GB}}$ (\ref{Gauss-Bonnet-Palatini}),
it is necessary
to see how the transformations (\ref{proj2}) and (\ref{weyl2}) act on the connections.
It is straightforward to see that the projective transformation changes both the trace
of the torsion and the non-metricity, but that the $B_\mu$ transformation only acts on
the trace of the torsion and leaves $Q_\mnr$ invariant:
\bea
&& T_\mn{}^\rho \ \rightarrow \ T_\mn{}^\rho
\ + \ 2(A_{[\mu} + B_{[\mu} ) \delta_{\nu]}^\rho ,
\nnw
&& Q_\mnr \ \rightarrow \  Q_\mnr \ + \ 2 A_\mu g_{\nu\rho}.
\label{AB-transformations}
\eea

There is a certain similarity, although also mayor differences, between our transformation
(\ref{weyl2}) and the torsion/non-metricity duality discussed in \cite{IPT}. There it was
shown that in $f(R)$ gravity the same physical situation can be described by different
geometrical descriptions, either in terms of the torsion or in terms of the non-metricity,
due to the fact that  the projective symmetry of these theories interchanges the degrees
of freedom of $T_{\mu\rho}{}^\rho$ and $Q_{\mu\rho}{}^\rho$ (see also \cite{BJJOSS} for a
similar observation in the context of the Einstein-Hilbert action). As can be seen from
(\ref{AB-transformations}), this property is not limited to four-dimensional $f(R)$ gravity,
but is present in any projectively invariant theory that allows the Weyl connection as a
solution. However, an important difference between our case and \cite{IPT} is that
the $B_\mu$ transformation in general is not a duality that relates physically equivalent
situations, but, as we will show, a solution generating transformation, that maps
certain connections onto other physically inequivalent ones.

As we mentioned before, the $B_\mu$ transformation is a symmetry when the theory is
restricted to the subset of metric-compatible connections, but not of the full theory.
This means that the connection space in the truncated theory $\mathcal{L}_{\mathrm{GB}}|_{Q=0}$ can
be divided into equivalence classes, which are the orbits of the $B_\mu$ transformations.
Two connections in the same orbit differ by the trace of the torsion and are physically
indistinguishable, as the $B_\mu$ transformation is a symmetry in $\mathcal{L}_{\mathrm{GB}}|_{Q=0}$.
Two connections in distinct orbits differ also in the traceless parts of the torsion.

However, from the point of view of the full theory $\mathcal{L}_{\mathrm{GB}}$, the $B_\mu$
transformation is not a symmetry, but a solution-generating transformation, as different
solutions of the (consistently)  truncated theory $\mathcal{L}_{\mathrm{GB}}|_{Q=0}$ are guaranteed
to be also solutions of the full theory. Within the $Q=0$ subset of the full theory,
the $B_\mu$ transformation hence maps solutions of the connection equation in other,
physically inequivalent solutions. On the other hand, outside the $Q=0$ subset, the
flow of the $B_\mu$ transformations also exists, but possibly map solutions of the theory
into connections that do not satisfy the equations of motion.

Finally, the projective transformation (\ref{proj2}) does not maintain solutions inside
the $Q=0$ subset, as it changes the trace of the non-metricity (as well as the trace of the
torsion). The orbits of the $A_\mu$ transformation that cross the $Q=0$ subset have a
pure-trace non-metricity, $Q_\mnr = \frac{1}{4} Q_{\mu\sigma}{}^\sigma g_{\nu\rho}$, while
the connections that have an additional non-trivial parts of $Q_\mnr$ lay on orbits of
$A_\mu$ that do not intersect the $Q=0$ subset. Since the projective transformation is
a symmetry of the full action $\mathcal{L}_{\mathrm{GB}}$, all connections on the same orbit of
$A_\mu$ are indistinguishable and hence physically equivalent.

\begin{figure}
\begin{center}
\leavevmode
\epsfxsize=9cm
\epsffile{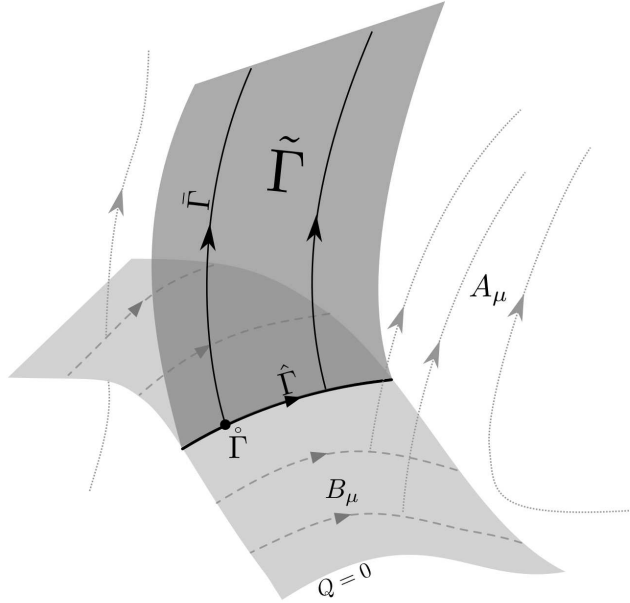}
\caption{\it The structure of the space of connections in four-dimensional
metric-affine Gauss-Bonnet theory: the $B_\mu$ transformations (\ref{weyl2}) act
as a solution-generating transformation in the subset of metric-compatible connections
($Q=0$), while the $A_\mu$ transformation relate physically equivalent connections, thanks
to projective symmetry of the theory. The solutions $\tilde \Gamma$ given in (\ref{GBsol})
form a subset spanned by the orbit of the $B_\mu$ transformation that contains the
Levi-Civita connection $\mathring{\Gamma}$ and the orbits of the  $A_\mu$ transformation
intersecting the aforementioned $B_\mu$ orbit. The orbits of  $A_\mu$ that do not cross
the $Q=0$ subset have a non-metricity tensor that is not pure trace,
$Q_\mnr \neq \frac{1}{4} Q_{\mu\sigma}{}^\sigma g_{\nu\rho}$.
}
\label{GBPimage}
\end{center}  
\end{figure}

With this structure in mind, we can see that the two-vector-family of solutions we have
found for the metric-affine Gauss-Bonnet action is of the general form
\be
\tilde \Gamma_\mn{}^\rho \ = \ \mathring{\Gamma}_\mn{}^\rho
 \ + \ A_\mu \, \delta_\nu^\rho \ + \ B_\nu \, \delta_\mu^\rho  
 \ - \ B^\rho \,g_\mn.
 \label{GBsol}
\ee
These solutions span a subset that is generated on the one hand by the $B_\mu$ orbit in
the $Q=0$ subset that contains the Levi-Civita connection and on the other hand by the
$A_\mu$ flow intersecting precisely this $\mathring{\Gamma}_\mn{}^\rho$ orbit (see Figure
\ref{GBPimage}). As far as we know, these are the only connections that are known to be
solutions to $\mathcal{L}_{\mathrm{GB}}$. But it should be clear that if a new solution
$\breve\Gamma_\mn{}^\rho$ were to be found on another one of the $B_\mu$ orbits in the
$Q=0$ subset, the flows of the $A_\mu$ and $B_\mu$ transformation would generate a new
two-vector-family  of solutions $\Gamma_\mn{}^\rho =  \breve{\Gamma}_\mn{}^\rho
+ A_\mu  \delta_\nu^\rho + B_\nu  \delta_\mu^\rho -  B^\rho g_\mn$. It seems therefore
reasonable to expect a (discrete or continuous) family of non-intersecting subset of
solutions, each one characterised by the orbits of the $B_\mu$ transformation that form
the intersection with the $Q=0$ plane.

We believe this structure not to be unique for the four-dimensional Gauss-Bonnet action, but
for any Lovelock theory in critical dimensions (\textit{i.e.}\ for the $k$-th order
Lovelock term in $D=2k$ dimensions). We believe the existence of the non-trivial solutions
is an indication of the non-topological character of Lovelock theories in critical
dimensions, in the presence of non-metric-compatible connections. 

On the other hand, not much is known about
the solutions of the four-dimensional Gauss-Bonnet action that are
not generated through the flows of the $A_\mu$ and $B_\mu$ transformations from the $Q=0$
subset, {\it i.e.}\ that have at least one part of the non-metricity that is not pure trace,
$Q_\mnr \neq \frac{1}{4} Q_{\mu\sigma}{}^\sigma g_{\nu\rho}$ (besides the general property that they
can be divided in the equivalence classes formed by the $A_\mu$ orbits). Similarly, to our
knowledge, there are no connections, other than (\ref{PalatiniGamma}), known to be a solution
of the Gauss-Bonnet action in dimensions $D>4$.

However, the fact that we have found non-trivial
(that is, non-equivalent) solutions for the specific four-dimensional case, disproves the
commonly accepted statement that the metric and the Palatini formalism are equivalent for
general Lovelock lagrangians. Indeed, even though the Levi-Civita connection is always a
solution to the metric-affine Lovelock actions, it is now clear that in general, higher-order
Lovelock theories can allow for physically distinct connections. It would be interesting to
find explicit non-trivial solutions for Lovelock theories in non-critical dimensions.

\section*{Acknowledgements}

The authors would like to thank Jos\'e Beltr\'an Jim\'enez, Tomi Koivisto,
Gonzalo Olmo, Miguel S\'anchez, Pablo S\'anchez-Moreno and
Jorge Zanelli for useful discussions.
This work was partially supported by the Spanish Ministry of Economy and
Competitiveness (FIS2016-78198-P), the Junta de Andaluc\'ia
(FQM101) and the Unidad de Excelencia UCE-PP2016-02 of the Universidad de Granada.
A.J.C. is supported by a PhD contract of the program FPU
2015 with reference FPU15/02864 (Spanish Ministry of Economy and Competitiveness) and
J.A.O. is supported by a PhD contract of the Plan Propio de la Universidad de Granada.
Our calculations have been checked by a computational program, making use of
xAct \cite{xact}.

\renewcommand{\theequation}{\Alph{section}.\arabic{equation}}
\appendix
\sect{$Q=0$ as a consistent truncation}
\label{Q=0}

In this Appendix we will show that metric-affine Lovelock theories restricted to the
subset of metric-compatible connections are consistent truncations of the full theories.
We will work in the
tangent space description, where the metric degrees of freedom are represented by the Vielbeins
$e^a{}_\mu$, which are the components of a local orthonormal coframe, and the affine connection
is substituted by the components of the connection one-form, $\omega_{\mu a}{}^b$, through the
appropriate basis transformation (sometimes  called the Vielbein  Postulate). The reason
is that the non-metricity of the connection in this set-up is given by the symmetric
part of the spin connection, $Q_{\mu}{}^{ab} = D_\mu \eta^{ab} = 2 \omega_\mu{}^{(ab)}$.

We start by considering a general action of the form
$S = \int \dder^Dx \,\cL(e, \mathcal{R}(\omega))$.
The variation of the action with respect to the affine connection, up to boundary terms
coming from partial integration, is given by
\bea
\delta_\omega S &=& \int \dder ^D x \, |e| \
\Sigma^{\mn a}{}_b \ \delta \mathcal{R}_{\mn a}{}^b(\omega)
\nnw
&=& -2\int \dder ^D x \, |e| \,
\Bigl[ (\nabla_\lambda - \half Q_{\lambda\sigma}{}^\sigma
       + \, T_{\lambda\sigma}{}^\sigma)\, \Sigma^{\lambda\mu a}{}_b
       \ - \ \half T_{\lambda\sigma}{}^\mu \, \Sigma^{\lambda\sigma a}{}_b\Bigr]
       (\delta \omega_{\mu a}{}^b),
\eea
where $\Sigma^{\mn a}{}_b = e^a{}_\alpha \, e^\beta{}_b \, \Sigma^{\mn\alpha}{}_\beta$, with
$\Sigma^{\mn\alpha}{}_\beta$ as in (\ref{Sigmadef}). The equation of motion, restricted to
metric-compatible affine connections is then of the form
\be
\Bigl(\tilde \nabla_\lambda
\,  + \, \tilde T_{\lambda\sigma}{}^\sigma \Bigr)\, \Sigma^{\lambda\mu a}{}_b\Bigr{|}_{Q=0}
\ - \ \half \tilde T_{\lambda\sigma}{}^\mu \, \Sigma^{\lambda\sigma a}{}_b\Bigr{|}_{Q=0}
\ = \ 0.
\label{EOMomega}
\ee
Here we used the notation $\tilde \omega_{\mu a}{}^b \equiv \omega_{\mu a}{}^b|_{Q=0}$ for
metric-compatible connections and $\tilde \nabla_\mu$ and $\tilde T_{\mn}{}^\sigma$ for
their covariant derivative and their torsion. Furthermore, $\Sigma^{\mn a}{}_b|_{Q=0}$ is
the $\Sigma^{\mn a}{}_b$-tensor (\ref{Sigmadef}), constrained to metric-compatible
connections, {\it i.e.}
\be
\Sigma^{\mu\nu\alpha}{}_\beta\Bigl{|}_{Q=0}\left.
\ = \ \frac{1}{\sqrtg} \, \frac{\delta S}{\delta \mathcal{R}_{\mu\nu\alpha}{}^\beta}\right|_{Q=0}.
\label{Sigmadef3}
\ee

On the other hand, consider now the the same theory
$S|_{Q=0} = \int \dder^Dx \,\cL(e, \tilde\mathcal{R}(\tilde\omega))$, but restricted to
connections that are metric-compatible, already at the level of the action. The variation
of this action with respect to the connection is then given by
\bea
\delta_{\tilde \omega} (S|_{Q=0}) &=& \int \dder ^D x \, |e| \
\tilde \Sigma^{\mn a}{}_b \ \delta \tilde \mathcal{R}_{\mn a}{}^b(\tilde\omega)
\nnw
&=& -2\int \dder ^D x \, |e| \,
\Bigl[ (\tilde \nabla_\lambda 
       + \, \tilde T_{\lambda\sigma}{}^\sigma)\, \tilde \Sigma^{\lambda\mu a}{}_b
       \ - \ \half \tilde T_{\lambda\sigma}{}^\mu \, \tilde \Sigma^{\lambda\sigma a}{}_b\Bigr] (
       \delta \tilde \omega_{\mu a}{}^b),
\eea
where now $\tilde \Sigma^{\mu\nu\alpha}{}_\beta$ is the $\Sigma$-tensor that arises from the
variation with respect to $\tilde \omega_{\mu a}{}^b$, 
\be
\tilde \Sigma^{\mu\nu\alpha}{}_\beta
\ = \ \frac{1}{\sqrtg} \, \frac{\delta S}{\delta \tilde \mathcal{R}_{\mu\nu\alpha}{}^\beta}.
\label{Sigmadef2}
\ee
The connection equation of $S|_{Q=0} = \int \dder^Dx \,\cL(e, \tilde\mathcal{R}(\tilde\omega))$ is
therefore of the form
\be
\Bigl(\tilde \nabla_\lambda 
       + \, \tilde T_{\lambda\sigma}{}^\sigma\Bigr)\, \tilde \Sigma^{\lambda\mu a}{}_b
       \ - \ \half \tilde T_{\lambda\sigma}{}^\mu \, \tilde \Sigma^{\lambda\sigma a}{}_b
\ = \ 0.
\label{EOMtildeomega}
\ee

In general, it turns out that
$\Sigma^{\mu\nu\alpha}{}_\beta|_{Q=0} \neq \tilde \Sigma^{\mu\nu\alpha}{}_\beta$. One way
of seeing this is by realising that  $\tilde \mathcal{R}_{\mu\nu\alpha}{}^\beta$ (and hence
also $\tilde \Sigma^{\mu\nu\alpha}{}_\beta$) is always antisymmetric in the last two indices,
but $\mathcal{R}_{\mu\nu\alpha}{}^\beta$ in general is not. Therefore, the connection equation
(\ref{EOMtildeomega}) of the truncated theory $S|_{Q=0}$ is in general not identical to the
truncated connection equation (\ref{EOMomega}) of the full theory $S$.

In fact, $S|_{Q=0}$ is a consistent truncation of $S$ if and only if the two $\Sigma$-tensors
coincide: $\Sigma^{\mu\nu\alpha}{}_\beta|_{Q=0} = \tilde \Sigma^{\mu\nu\alpha}{}_\beta$.
In particular, this turns out to be the case for the Gauss-Bonnet action
(\ref{Gauss-Bonnet-Palatini}) and, more generally, for all metric-affine Lovelock theories
(\ref{Lovelock-Palatini}). Indeed, from (\ref{LL-Sigma}) it is straightforward to see that
\bea
\tilde  \Sigma^{\mu\nu\alpha}{}_\beta
&=& k \ \delta^{\mu\nu\rho_2\lambda_2 \dots \rho_k\lambda_k}_{\alpha\beta\gamma_2\epsilon_2 \dots \gamma_k\epsilon_k} \
\tilde \mathcal{R}_{\rho_2\lambda_2}{}^{\gamma_2\epsilon_2} \, \dots \,
\tilde \mathcal{R}_{\rho_k\lambda_k}{}^{\gamma_k\epsilon_k} \nnw
&=& k \ \delta^{\mu\nu\rho_2\lambda_2 \dots \rho_k\lambda_k}_{\alpha\beta\gamma_2\epsilon_2 \dots \gamma_k\epsilon_k} \
\mathcal{R}_{\rho_2\lambda_2}{}^{\gamma_2\epsilon_2}\Bigl{|}_{Q=0} \, \dots \,
\mathcal{R}_{\rho_k\lambda_k}{}^{\gamma_k\epsilon_k}\Bigl{|}_{Q=0}  \nnw
&=& \Sigma^{\mu\nu\alpha}{}_\beta\Bigl{|}_{Q=0},
\eea
since by definition
$\tilde \mathcal{R}_{\mu\nu\alpha}{}^\beta = \mathcal{R}_{\mu\nu\alpha}{}^\beta|_{Q=0}$.


\end{document}